\documentclass{iopart}
\usepackage{iopams}

\usepackage{graphicx}
\usepackage{latexsym}

\newcommand{\symmord}[1]{\boldsymbol{:}\!\!{#1}\!\!\boldsymbol{:}}

\def\be {\begin{equation}}
\def\ee  {\end{equation}}
\def\bea {\begin{eqnarray}}
\def\eea {\end{eqnarray}}

\newcommand{\eqref}[1]{(\ref{#1})}
\newcommand{\Hil}{\mathcal{H}}
\newcommand{\re}{\mathbb{R}}
\newcommand{\rd}{{\rm d}}
\newcommand{\rv}{{\rm v}}
\newcommand{\lPl}{\ell_{\rm Pl}}
\newcommand{\N}{\mathcal{N}}

\newcommand{\natu}{\mathbb{N}}
\newcommand{\Sp}{{\rm Sp}}

\newcommand{\Sch}{\mathcal{S}}

\newcommand{\id}{\mathbb{I}}
\newcommand{\ket}[1]{{|#1\rangle}}
\newcommand{\ub}[1]{\underline{#1}}

\newcommand{\Ham}{\boldsymbol{H}}
\newcommand{\expx}[1]{\left\langle #1 \right\rangle}
\newcommand{\exps}[1]{\langle #1 \rangle}

\begin{document}

\title{Dust reference frame in quantum cosmology}

\author{Viqar Husain and Tomasz Paw{\l}owski}

\address{Department of Mathematics and Statistics\\University of New Brunswick\\
  Fredericton, NB E3B 5A3, Canada}

\ead{vhusain@unb.ca, tpawlows@unb.ca}

\begin{abstract}

We give a formulation of quantum cosmology with a pressureless dust
and arbitrary additional matter fields.  The system has the property that its
Hamiltonian constraint is linear in the dust momentum. This feature provides a
natural time gauge, leading to a physical hamiltonian that is not a square root.
Quantization leads to  Schr{\"o}dinger equation for which unitary evolution is directly
linked to geodesic completeness. Our approach simplifies the analysis of both
Wheeler-deWitt and loop quantum cosmology (LQC) models, and significantly broadens
the applicability of the latter. This is demonstrated for arbitrary scalar field
potential and cosmological constant in LQC.

\end{abstract}


\section{Introduction}

In attempts to formulate a  non-perturbative theory of  quantum gravity one of the
obstacles is the so-called problem of time. Its manifestation in the canonical formulation
of general relativity is that the Hamiltonian vanishes, which is a direct consequence of
the time reparametrization invariance of the theory. For matter with the usual  kinetic
energy, the Hamiltonian constraint is quadratic in all momenta, including the pure gravity
term.

There appears to be an obvious way to address this problem: impose an explicit  time gauge
fixing to obtain a true Hamiltonian that describes the evolution of  physical degrees of
freedom.  This explicit  deparametrization brings with it some difficulties: the true
Hamiltonian is generically  a square root, which is difficult to  define as an operator. A
second problem is that different time gauge fixings give radically different true
Hamiltonians, so in the final analysis the quantum theory ends up being  time-gauge
dependent.

In a recent paper \cite{hp1} we presented a complete non-perturbative quantization of
gravity that solves both problems. This approach uses a single timelike dust field
(motivated by the Brown and Kucha\v{r} (BK) form)  \cite{brown-kuchar} coupled to general
relativity and standard model matter. Like the BK action, the canonical theory has the
feature that the Hamiltonian constraint is linear in the dust momentum. However, unlike
the BK action, it also has the remarkable feature that in a natural time gauge fixing {\it
the true Hamiltonian is not a square root.} Furthermore, this Hamiltonian is identical in
form to what would be the Hamiltonian constraint without the dust field. It is this
feature, combined with the well-developed kinematical framework for loop quantum gravity,
that leads to a complete quantum theory.

In this paper we  apply this development to quantum cosmology in both the Wheeler-DeWitt 
and LQG  quantization. We show that  (i)  the true Hamiltonian is first order,  and in the LQC case it
poses no self-adjoint extension problem, (ii) the non-zero cosmological constant is analytically
solvable for both signs, and (iii) any type of matter is included.

Our approach differs significantly from existing LQC models, all of which  use the scalar
field as a clock;  this relies on the scalar field momentum being a constant of  motion, and 
leads to a Hamiltonian constraint operator that is second order in
time (and so is not a time-dependent Schr{\"o}dinger equation).  For these reasons  it is difficult to
implement scalar field time for arbitrary scalar potential, including polynomial or slow
roll cases used in inflationary models.

\section{General relativity  with dust}

We consider the class of the flat isotropic spacetimes whose metric is  of the form
\begin{equation}\label{eq:FRW-metric}
  \rd s^2\ =\ -N^2 \rd t^2 + a^2(t)(\rd {\rm x}^2 + \rd {\rm y}^2 +\rd {\rm z}^2) \ .
\end{equation}
The theory is given by the Einstein-Hilbert action with timelike dust field $T$
and  an arbitrary matter lagrangian ${\cal L}_m$ (assumed to be at most second order in time):
\begin{equation}
  \eqalign{
    S\ &=\ \frac{1}{4G} \int \rd^4x \sqrt{-g} R\ 
	-   \int \rd^4x \sqrt{-g}\  {\cal L}_m  \cr
    &\hphantom{=}\ - \frac{1}{2} \int  \rd^4 x\  \sqrt{-g} M ( g^{ab}\partial_a T \partial_b T + 1) \ . \cr
  }
\end{equation}
where $M$ is a Lagrange multiplier. The last term resembles the Brown-Kucha\v{r} action \cite{brown-kuchar}, but contains only one scalar.

The canonical formulation of this action gives a Hamiltonian constraint
linear in the dust momentum $p_T$ conjugate to $T$, together with the usual spatial diffeomorphism
constraint \cite{hp1}. The linearity in $p_T$ makes the theory resemble a parametrized system, and 
suggests the gauge $T=t$.  In this gauge the true Hamiltonian is $-p_T$, given by
\begin{equation}
  \Ham\ \equiv\ -p_T = \Ham_G + \Ham_m\,
\end{equation}
where $\Ham_G$ and $\Ham_m$ are the gravitational and matter parts of the Hamiltonian constraint
coming from the first two terms in the action. In this natural gauge, quantization gives
the time dependent Schr\"odinger equation
\begin{equation}\label{eq:Sch-cosm}
  i\frac{\partial \Psi}{\partial t}\
  =\ \left( \hat{\Ham}_G + \hat{\Ham}_m \right) \Psi
\end{equation}
Since $\Ham_G$ and $\Ham_m$ are manifestly independent of the dust variables, the physical
hamiltonian is time independent. The spatial diffeomorphism constraint of course vanishes identically
in the homogeneous sector.

The dust effectively provides a cosmic time, therefore there is a direct relation between
unitarity and geodesic completeness of the resulting spacetime in this time parameter;
if classical singularities are not removed by quantum dynamics, loss of
unitarity occurs precisely at the singular points. In the following we will see illustrative
examples of this.

\section{Quantum cosmology with dust}\label{ssec:L0}

As a first example we consider FRW cosmology with only dust, ie. $\Ham_m =0$ in
\eqref{eq:Sch-cosm}, and give a quantization  of this system using both the Wheeler-DeWitt
(WDW)  and the background independent LQG method. These methods differ in the choice of
Hilbert space used. This will demonstrate a first advantage over scalar field time used in
all previous LQC work: classically the singularity is reached in \emph{finite} dust time,
so it  allows a direct comparison of the mechanism of singularity resolution in background
independent quantization,  and the lack of it in WDW quantization.

The phase space variables are the canonical pair $\{a,P_a\}=1$ where $a$ is 
the scale factor \eqref{eq:FRW-metric}. Classically $a$ is positive, therefore  to have   
phase space topology $\re^2$\footnote{This gives a simple transformation between the 
  configuration and momentum representations.
} one needs to equip $a$ with an orientation. 
To do this  we can either work with the original variables or use a canonical transformation
to the oriented volume $v$ ($= \alpha^{-1} a^3$) \footnote{  $\alpha\approx 1.35\lPl^3$. 
  This choice synchronizes the variable $v$ with what is used in LQC. $\alpha$ is expressed 
  in terms of the Barbero-Immirzi parameter and the LQC area gap \cite{aps-imp}. 
} and its conjugate momentum $b$, 
which satisfy the Poisson bracket $\{v,b\}=2$. In these variables $\Ham_G = (3\pi %
G/2\alpha) b^2|v|$.  

\subsection{Wheeler-DeWitt quantization}

In WDW quantization one uses the standard Schr\"odinger representation. The canonical 
variables $(v,b)$ are promoted to operators defined on the Schwartz space 
$\Sch(\re)\in \Hil_G = L^2_s(\re,\rd v)$, where $s$ denotes restriction to the 
functions symmetric in $v$. Since the classical Hamiltonian is linear in $v$,  a 
naive realization of the Hamiltonian operator in momentum space  is  
\begin{equation}
\label{eq:WDW-Ham-b}
 \tilde{\Ham}_G\ := \  3i\pi\lPl^2\alpha^{-1}\, \hat{b}^2\partial_b  \ .
\end{equation}
It acts on a dense  domain  $\tilde{\Hil}_G$ of functions spanned by the eigenfunctions
$\theta(\pm x)\exp(ikx)$, where $k\in\re$ and $x=-1/b$. This operator however breaks
the positive definiteness of the classical Hamiltonian since its spectrum
$\Sp(\tilde{\Ham}_G)=\re$.  An obvious way to fix this problem is just to restrict 
attention to the positive part of \eqref{eq:WDW-Ham-b}.   
However there is a rigorous way
to  arrive at this  conclusion in the following steps: 
(i) Begin with the the natural factor ordering \cite{aps-imp,klp-gave} 
\begin{equation}\label{eq:WDW-Ham}
  \hat{\Ham}_G\ =\ \frac{3\pi G}{2\alpha} \sqrt{|\hat{v}|}\,\hat{b}^2\,\sqrt{|\hat{v}|}.
\end{equation}
This is a symmetric positive definite operator on $\Sch(\re)$, 
(ii) analysis of its deficiency subspaces shows that it  admits a family of extensions 
labeled by $e^{i\beta}\in U(1)$, and that  the spectrum of each extension $\hat{\Ham}_{\beta}$ 
with domain  $\Hil_\beta$ is \emph{nondegenerate}. 
(iii) Due to this fact, for each extension $\hat{\Ham}_{\beta}$, there is an invertible  map
$\ub{P}_{\beta}:\Hil_{\beta}\to\tilde{\Hil}_{\beta}$   to a certain \emph{auxiliary space} 
such that $\hat{\Ham}_G$ is mapped onto the positive part $[\tilde{\Ham}_G]^+$ of $\tilde{\Ham}_G$. 
(The auxiliary spaces $\tilde{\Hil}_{\beta}$ are determined by the selfadjoint  extensions 
$\tilde{\Ham}_{\beta}$ of $\tilde{\Ham}_G$, and construction of the map uses the fact 
that the symmetric state is encoded in  one orientation  $\theta(v)\Psi(v)$.)   

In each extension the  physical states are 
\begin{equation}\label{eq:WDW-state}
  [\ub{P}_{\beta}\Psi](x)\ =\ \int_{\re^+} \rd k\, \tilde{\Psi}(k)\,
  [\theta(x) e^{ikx} + \theta(-x) e^{i\beta} e^{ikx}] \ ,
\end{equation}
where $\tilde{\Psi}\in L^2(\re,k\rd k)$, with the inner product 
\begin{equation}\label{eq:WDW-IP-b}
  \langle \ub{P}_{\beta}\Psi|\ub{P}_{\beta}\Phi \rangle\ =\
  \int_{\re} \rd x \overline{[\ub{P}_{\beta}\Psi]}(x)[i\partial_x][\ub{P}_{\beta}\Phi](x) \ .
\end{equation}
 Time evolution is given by
\begin{equation}\label{eq:WDW-evol-map}
  \tilde{\Psi}_{t}(k)\ = e^{i\omega(k)(t-t_o)} \tilde{\Psi}_{t_o}(k) \ , \quad
  \omega(k) = 3\pi\lPl^2\alpha^{-1} k \ ,
\end{equation}
 This is  a plane wave packet moving with the constant speed $\rv:= 3\pi\lPl^2\alpha^{-1}$
till it hits $x=0$, where it gets rotated by an extension dependent phase $e^{i\beta}$.

The dynamics of the system is described by the observable $\hat{V}=|\hat{a}^3|$.
Under the mapping $\tilde{V}=\ub{P}_{\beta}\hat{V}\ub{P}_{\beta}^{-1}$, it takes a simple form
involving only $\partial_x$ and $x^2\id$. Its time evolution and
dispersion is analytically determined:  \footnote{
  The explicit dependence of the  first term on $\alpha$ is compensated by analogous dependence
  in $\exps{\hat{k}}$.
}
\begin{equation}\label{eq:WDW-v-evol}
 \exps{\hat{V}}(t)= V(t)\ =\ 2\alpha^{-1} \exps{\hat{k}}\  [3\pi\lPl^2(t-t_s)]^2 + 2\alpha\tilde{\sigma}_x^2\ ,
 \ \ \ \hat{k} = k\id\ , 
\end{equation}
where $t_s$ corresponds to the (always existing) point where $\exps{\symmord{i\partial_x \hat{x}}}=0$
(the colons denote symmetrization) and $\tilde{\sigma}_x^2$ can be related with the dispersion 
of $x$ at this moment. In the same way one can determine the time dependence of Hubble parameter $H$.
Physically these trajectories follow the classical ones, and describe a contracting universe
until  the big crunch singularity at the time $t_s$. At that moment $V$ becomes comparable
to its dispersion. In this sense the singularity $V=0$ is \emph{reached dynamically}.

To determine the evolution past the singular point one needs to know the selfadjoint
extension. This is equivalent to imposing additional boundary conditions at the singularity
$x=0$; it cannot be determined by an analysis of the state for $t<t_s$. Given a choice of 
extension it is possible to continue the evolution. The resulting trajectory for $t>t_s$ 
describes  a universe that expands out of the singularity and subsequently follows a classical 
trajectory. Thus we see that in WDW
quantization the big bang singularity is dynamically reached, and evolution beyond it is
not uniquely determined without additional boundary data. In this sense the the WDW quantization
\emph{does not resolve} the singularity.

\subsection{Loop Quantum Cosmology}
We now consider the same problem  in LQC. To quantize the system we follow the
improved dynamics prescription of \cite{aps-imp}. As in the WDW case,
we can apply directly the results of  kinematical quantization  \cite{aps-imp,acs}. The canonical 
variables are  again $(v,b)$,  but the physical Hilbert space is now 
$\Hil = L^2(\bar{\re},\rd\mu_B)$, where $\bar{\re}$ is the Bohr compactification of
the reals and $\rd\mu_B$ is the Haar measure on it. The (physical) gravitational Hamiltonian
$\Ham_G$ is a second order difference operator in $v$. With the same factor
ordering as in  \eqref{eq:WDW-Ham}, it takes the form
\begin{equation}\label{eq:HG-L0-1}
  \hat{\Ham}_{G}\ =\ -\frac{3\pi G}{8\alpha} \sqrt{|\hat{v}|}
    (\hat{\N}-\hat{\N}^{-1})^2 \sqrt{|\hat{v}|} \ , \quad
  \hat{\N}\ket{v}\ =\ \ket{v+1} \ .
\end{equation}
This operator is non-negative definite and, unlike in WDW quantization, it is \emph{essentially 
selfadjoint} \cite{kl-flat}. The fact, that $\hat{\Ham}_{G}$ does not couple the sets $v<0$, 
$v=0$ and $v>0$ permits with construction of the mapping $P:\Hil_G\to\tilde{\Hil}_G$ into an auxiliary
space analogous to that for WDW quantization. Due to the selfadjointness of
$\hat{\Ham}_{G}$ its construction here is even simpler.
The auxiliary Hamiltonian now takes the form
\begin{equation}
  \tilde{\Ham}_G\ :=\ P \hat{\Ham}_G P^{-1}\ =\ [ 3i\pi\lPl^2\alpha^{-1}\sin^2(b)\partial_b ]^+ \ ,
\end{equation}
where due to discreteness in $v$ the momentum $b$ is periodic with period $\pi$.
The auxiliary Hilbert space $\tilde{\Hil}_G$ consists of states
\begin{equation}\label{eq:L0-state}
  [P\Psi](x)\ =\ \int_{\re^+} \rd k\ \tilde{\Psi}(k) e^{-ikx} \ , \quad
  \tilde{\Psi} \in L^2(\re^+,k\rd k) \ ,
\end{equation}
where now $x=-\cot(b)$. The inner product (in new coordinate) is given by the same
formula \eqref{eq:WDW-IP-b}.
At this point it is worth noting that $\tilde{\Ham}_G$ is a $1$st order differential
operator so its spectrum is nondegenerate. This fact is crucial in identifying
the images of the mapping of the original Hilbert space basis to the auxiliary one.
In the models with a scalar field an analogous construction would lead to the $2$nd
order operator with degenerate spectrum. This make the identification of the bases
much more involved and required an alternative approach \cite{acs} (which
cannot be extended to more complicated systems.)

The physical evolution is again given by \eqref{eq:WDW-evol-map}, but  due to selfadjointness
the evolution across $x=0$ is now unique. In particular the expectation value of the volume is now 
\begin{equation}
  V(t)\ =\ 2\alpha^{-1} \exps{\hat{k}}\  [3\pi\lPl^2(t-t_s)]^2
  + 2\alpha\tilde{\sigma}_x^2 + 2 \alpha \exps{\hat{k}} \ ,
\end{equation}
where the meaning of $t_s$ and $\tilde{\sigma}_x^2$ is the same as in \eqref{eq:WDW-v-evol}.
We note here the presence of a nontrivial minimum volume $V_m = 2\alpha \exps{\hat{k}}$, which  
implies  dynamical resolution of the singularity through the bounce, as
in the models with scalar field time \cite{aps-imp,acs}.

In summary we have seen  that the  dust reference frame gives exactly solvable models 
in WDW and LQC,  a feature present in earlier models  \cite{acs}.  We now turn to 
more complicated models, where the real advantage of dust time  becomes
apparent. 

\subsection{LQC with a cosmological constant}

FRW cosmologies with non-vanishing cosmological constant and scalar field time have been
investigated in detail in LQC  \cite{aps-imp,bp-negL,ap-posL}. In these works  the
appearance of the square root in the evolution operator rendered the models analytically 
unsolvable. Furthermore, for  the case $\Lambda>0$ the Hamiltonian was not
essentially selfadjoint; the multitude of extensions complicated the analysis even
further, in particular raising an issue of dependence of global evolution on
the choice of time \cite{klp-gave}. As we will see, the situation simplifies considerably
with dust time.

With the cosmological constant the gravitational Hamiltonian is 
\begin{equation}
  \Ham_{\Lambda} = \Ham_G + \frac{3\rho_c}{16\alpha}\Lambda|v| \ ,
\end{equation}
(where $\rho_c\approx 0.81\rho_{\rm Pl}$ is the critical energy density of LQC
\cite{aps-imp}.) The polymer quantization of this operator is straightforward \cite{aps-imp,bp-negL}:
It is positive definite for $\Lambda<0$ and \emph{essentially selfadjoint} for \emph{all values of $\Lambda$}.
Remarkably, the cosmological constant term does not disrupt the properties required 
for the mapping  $P_{\Lambda}:\Hil_G\to\tilde{\Hil}_{\Lambda}$ between the physical and 
auxiliary Hilbert spaces.  The auxiliary Hamiltonian is  
\begin{equation}\label{eq:Ham-aux-L}
  \tilde{\Ham}_{\Lambda}\ :=\ P_{\Lambda}\hat{\Ham}_{\Lambda}P_{\Lambda}^{-1}\
  =\ \tilde{\Ham}_{G} - \frac{3i\rho_c}{8\alpha}\Lambda\partial_b \ ,
\end{equation}
This is simple enough  to determine its eigenfunctions analytically. For all
values of $\Lambda$ these can be expressed in the form $e^{ikx(b)}$, where 
 $x(b)$ depends analytically on  $\Lambda$.

\noindent \underbar{$\Lambda<0$:} For this case we have
\begin{equation}
  x(b)\ =\ \arctan \left[ \sqrt{(3\rho_c|\Lambda|+24\pi\lPl^2)/(3\rho_c|\Lambda|)} %
 \  \tan(b) \right]\ ,
\end{equation}
This inherits the periodicity properties of $b$. Therefore the spectrum
of the auxiliary Hamiltonian $[\tilde{\Ham}_{\Lambda}]^+$ is discrete. Nondegeneracy
of its spectrum allows again to relate the original Hilbert space basis elements
with the auxiliary counterparts in a straightforward way. The (auxiliary) physical
states are
\begin{equation}\label{eq:state-negL}
  \tilde{\Hil}_{\Lambda} \ni [P_{\Lambda}\Psi](x)\ =\ \sum_{n\in\natu} \tilde{\Psi}_n e^{2inx} \ , \quad
\end{equation}
and the time evolution is given by 
\begin{equation}
  \tilde{\Psi}_n(t) =  \tilde{\Psi}_n(t_o) e^{i\omega_n (t-t_o)} \ , \quad
  \omega_n = \frac{n}{4\alpha} \sqrt{3\rho_c|\Lambda|(3\rho_c|\Lambda|+24\pi\lPl^2)}  \ .
\end{equation}
The standard observables used for $\Lambda=0$ also map in a straightforward way into
an auxiliary system. In particular  $\tilde{V} :=
P_{\Lambda}\hat{V}P_{\Lambda}^{-1}$ of $\hat{V}$ acts  directly  
on the spectral profile $\tilde{\Psi}_n$, in terms of the shift and multiplication
operators.

The calculation of expectation values reproduces
the dynamics qualitatively similar to the one of the scalar field system: the universe
goes through an infinite chain of quantum bounces and classical recollapses.
Now however, unlike in \cite{bp-negL} due to exact uniformity of $\Sp(\hat{\Ham}_{\Lambda})$
the evolution is \emph{exactly periodic}, thus the wave packets \emph{do not spread out}
between the cycles of the evolution.

\noindent \underbar{$\Lambda>0$:}  As in the model with scalar field we observe qualitative changes
at the critical value $\Lambda_c=8\pi G\rho_c$. The most interesting case is for 
$0<\Lambda<\Lambda_c$. Let us write $\Lambda$ as $\Lambda=\Lambda_c\sin^2(\beta)$, with $\beta\in [0,\pi/2]$.  
Then  the domain of $b$ ($[0,\pi]$) admits two uncoupled deSitter sectors.
Furthermore the spectrum of auxiliary Hamiltonian \eqref{eq:Ham-aux-L} is degenerate,
which makes the identification of the basis vectors of $\tilde{\Hil}_{\Lambda}$ considerably
more involved. Fortunately a careful analysis of the asymptotic properties of the eigenfunctions
of $\tilde{\Ham}_{\Lambda}$ allows them to be uniquely  identified. This gives 
that $\Sp(\tilde{\Ham}_{\Lambda})=\re$ with the eigenfunctions 
\begin{equation}\label{eq:posL-bas}
  \eqalign{
    e_{+|k|}(b)\ &=\ \theta(\sin(b)-\sin(\beta)) e^{ix_+(b)}  \ , \cr
    e_{-|k|}(b)\ &=\ \theta(\sin(\beta)-\sin(b)) e^{ix_-(b)} \ ,
  }
\end{equation}
where the coordinates $x_{\pm}$ are related to $b$ by
\begin{equation}
  \tan(\beta) [\tanh(x_{\pm})]^{\mp 1}\ =\ \tan(b) \ , \quad x_{\pm}\in\re \ .
\end{equation}
We have thus two sectors, one of positive and the other of negative energy, supported respectively 
on $\sin(b)>\sin(\beta)$ and $\sin(b)<\sin(\beta)$. In each sector, physical states 
are of a form analogous to \eqref{eq:L0-state},  with $k\in\re$\footnote{Each sector 
  is represented by $k>0$ and $k<0$ respectively.}, a basis defined via \eqref{eq:posL-bas} and 
$\tilde{\Psi}\in L^2(\re,|k|\rd k)$. They evolve as
free plane wave packets in the respective coordinates $x_{\pm}$. Time evolution is again
given by eqn. \eqref{eq:WDW-evol-map} , now with the frequencies
$\omega(k) = 3\pi\lPl^2\alpha^{-1}\cos^2(\beta) k$.
The volume operator  is in this case  involves 
$\partial_{x_{\pm}}$ and multiplications by hyperbolic functions of $x_{\pm}$. One can
thus apply the methods of \cite{acs} to calculate its expectation value and dispersion analytically.

The positive energy sector  gives {\it analytical} results similar to those found with scalar 
field time in LQC, ie.  a contracting  deSitter-like  universe bouncing once into an expanding  one.
In addition we obtain  negative energy solutions  which give DeSitter universes with ``phantom" 
dust and classical bounce, i.e. due to classical negative energy.  The last feature has interesting  
implications for the case  $\Lambda\geq\Lambda_c$:   there is a large solution space of phantom 
dust solutions with classical bounce ($\Lambda=\Lambda_c$) or cyclic evolutions consisting 
of an infinite chain of classical bounces and quantum recollapses ($\Lambda>\Lambda_c$). 
These cases however are just mathematical curiosity and are not of physical interest.

\section{Other applications}

As we have seen, the use of dust time considerably simplifies the analysis
of systems with cosmological constant. Its usefulness is not however restricted to only 
these: {\it the physical Hamiltonian is self-adjoint for any non-exotic matter}, including
Yang-Mills gauge fields, and scalar fields with any potential term. This follows from
the fact that $\Ham_G$ is self-adjoint for any value of the cosmological  constant
\cite{klp-aspects}, and that the only gravitational variables the matter terms contain are
factors of the volume operator and its inverse. This holds also for other spatial topologies.

\underbar{\it Inflaton potential:} One of potentially important applications is the analysis of the cosmologies
with massive scalar field, including those with inflaton potential. In this situation
the phase space is coordinatized by $(v,b,\phi,p_{\phi})$, where $\phi$ is the scalar
field value and $p_{\phi}$ its momentum. The physical Hilbert space is simply a product
\begin{equation}
  \Hil\ =\ \Hil_G \otimes \Hil_{\phi} \ , \quad \Hil_\phi\ =\ L^2(\re,\rd\phi) \ ,
\end{equation}
and the Hamiltonian is $\hat{\Ham}_G + \hat{\Ham}_{\phi}$, where $\hat{\Ham}_{\phi}$ is  
a polymer quantized scalar field Hamiltonian \cite{phi-poly}. The physical evolution is
given by the Schr{\"o}dinger equation \eqref{eq:Sch-cosm} and  physical
observables can be taken as the kinematical ones of \cite{aps-imp}, say
$\hat{V}, \hat{\phi}$. Furthermore the mappings into auxiliary systems found for
$\hat{\Ham}_{\Lambda}$ can be extended to this situation, thus casting the physical
system into one that is at least numerically  manageable.

\underbar{\it Effective Friedman equation:} Another application is a derivation of the
effective Friedmann equation. Here we have the advantage that  LQC kinematical operators become
\emph{physical} ones. Given a physical state, the derivation
requires computation of the expectation values of the square of the Hubble operator and of the physical
energy density operator of the dust. The latter is given by the operator $\hat{\rho}_D:= -:\!\!V^{-1} \hat{\Ham}_G\!:$
 and  the inverse volume operator has a suitable definition that descends from LQG.
In a chosen factor ordering, the Hubble operator $\hat{H}$ and matter density $\hat{\rho}$ have the simple forms
\begin{equation}
  \hat{H}\ =\ \pi\lPl^2\alpha^{-1} \sin(2\hat{b}) \ , \quad
  \hat{\rho}\ =\ \rho_c\sin^2(\hat{b}) \ .
\end{equation}
This allows us  to express $\hat{H}^2$ in terms of $\hat{\rho}$. Taking the expectation values
one arrives to the modified Friedmann equation
\begin{equation}\label{eq:Fried-eff}
  \expx{\hat{H}}^2\
  =\ \frac{8\pi G}{3} \expx{\hat{\rho}_D} \left( 1 - \frac{\expx{\hat{\rho}_D}}{\rho_c} \right)\
    -\ \left[ \frac{8\pi G}{3} \frac{\sigma_\rho^2}{\rho_c} + \sigma_H^2 \right] \ ,
\end{equation}
where $\sigma_{H}$, $\sigma_{\rho}$ are the dispersions of $\hat{H}$ and $\hat{\rho}$
respectively.

This equation is exact  for all elements of $\Hil_G$ and generalizes to \emph{any} matter type. 
For if the term $\Ham_{\phi}$ in \eqref{eq:Sch-cosm} is nonzero, the above equation remains valid: 
the formula for energy density is unchanged, but the density is now to be interpreted as that 
of all non-gravitational matter: $\rho_{\rm tot}:=- \symmord{V^{-1}\Ham_G}$.
The form of \eqref{eq:Fried-eff} implies a bounce for all  matter for which  energy density 
grows unboundedly with $V^{-1}$.

\underbar{\it Classical effective dynamics:}   
Our dust frame formalism  can also   be applied to  the ``effective'' approach  developed 
in \cite{b-eff}, where it  suggests a useful  improvement. There one begins with a pair of 
canonical variables $(x,p)$ (or set of pairs) subject to constraints, and defines quantum 
evolution via a  set of equations of motion for the quantities $\exps{\hat{x}}$, $\exps{\hat{p}}$ 
and $G^{m,n}=\exps{(\hat{x}-\exps{\hat{x}}\id)^m (\hat{p}-\exps{\hat{p}}\id)^n}$. 
Provided the  constrained system is under sufficient control, the equations for expectation 
values can describe to a good precision the dynamics for sufficiently large classes of physical states. 
However implementation of this approach requires a procedure for converting kinematical 
observables to the physical ones. Explicit deparametrization  produces a square root problem  
which requires careful approximation, and this has not yet been satisfactorily achieved. 
In the dust frame presented here, this problem vanishes allowing this approach to be unambiguously implemented.  

\section{Summary}

We have  presented applications to cosmology of a formulation of quantum gravity
with dust time in which the gravitational part of the Hamiltonian constraint becomes
the true physical Hamiltonian.  This has several consequences which follow directly
from the fact that the kinematical Hilbert space and observables become physical.
Foremost among them is that the physical Hamiltonian is not a square root and so 
allows an analytical treatment in cases where it was not possible before, and permits
a clear numerical approach  where necessary. 

We studied several examples in the WDW and LQC with cosmological constant,
all of which were shown to be  analytically solvable. We also outlined possible further developments,
including extensions to any matter type and potential, which are not possible with scalar field time.

\ack
This work was supported by the Natural Science and Engineering Research Council of Canada.

\section*{References}

\bibliographystyle{iopart-num}
\bibliography{hp-dustLQC-FTC}

\end{document}